\documentclass[aps, amsfonts, amsmath, nofootinbib, showpacs, byrevtex, 
preprintnumbers, showkeys, prd]{revtex4}   
\usepackage{graphicx}
\usepackage{bbm}

\newcommand{\bi}{\bigskip}
\newcommand{\no}{\noindent}
\newcommand{\be}{\begin{equation}}
\newcommand{\ee}{\end{equation}}
\newcommand{\bea}{\begin{eqnarray}}
\newcommand{\eea}{\end{eqnarray}}
\newcommand{\hk}{\hspace{0.1cm}}
\newcommand{\hs}{\hspace{0.5cm}}

\newcommand{\rk}{\right)}
\newcommand{\lk}{\left(}

\newcommand{\sli}{\sum\limits}

\newcommand{\il}{\int\limits}

\newcommand{\vA}{\vec{A}}
\newcommand{\vx}{\vec{x}}
\newcommand{\vy}{\vec{y}}

\newcommand{\Id}{ \mathbbm{1} }

\renewcommand{\vec}[1]{\mbox{\boldmath$#1$\unboldmath}}

\begin{document}

\title{The Wilson loop from a Dyson equation}

\author{M. Pak and H. Reinhardt}
\affiliation{Institut f\"ur Theoretische Physik\\
Auf der Morgenstelle 14\\
D-72076 T\"ubingen\\
Germany}
%



%
%
%
\begin{abstract}
The Dyson equation proposed for planar temporal Wilson loops in the context of
supersymmetric gauge theories is critically analysed thereby exhibiting its
ingredients and approximations involved. We reveal its limitations and
identify its range of applicability in non-supersymmetric gauge theories.
In particular, we show that this equation is applicable only to strongly
asymmetric planar Wilson loops (consisting of a long and a short pair of loop
segments) and as a consequence the Wilsonian potential can be
extracted only up to intermediate distances.
By this equation the Wilson loop is exclusively determined by the gluon
propagator. We solve the Dyson equation in Coulomb gauge 
for the temporal Wilson loop with the instantaneous part of the gluon propagator and 
for the spatial Wilson loop with the static gluon
propagator obtained in the Hamiltonian  
approach to continuum Yang-Mills theory and on the lattice. In both cases we find a linearly rising color
potential.
\end{abstract}

\maketitle
\bi

\no
\section{Introduction}
The Wilson loop 
is a quantity of central interest in Yang-Mills theory. The
temporal Wilson loop is related to the static potential between infinitely heavy
color sources and represents the order parameter of confinement: An area law
in the temporal Wilson loop corresponds to a linearly rising potential.
Furthermore the potential $V(L)$ extracted from a rectangular temporal Wilson
loop of spatial extension $L$ is the ground state energy of a pair of an
infinitely heavy quark and antiquark at separation $L$. As discussed in detail
in Ref.~\cite{Cucchieri:2000hv} 
this ``Wilsonian'' potential $V(L)$ is not the
fundamental color confining potential $V_C (L)$, but rather the residual
potential that survives color screening by vacuum polarization effects, i.e.~$V(L)$ contrary to $V_C (L)$ contains the back reaction of the Yang-Mills
vacuum to the presence of the infinitely heavy color sources. Obviously, the
screening of color charges will lower the energy so that $V(L) < V_C (L)$.
Therefore, a linearly rising fundamental potential $V_C (L)$ does not necessarily imply
an area law in the corresponding Wilson loop.
The fundamental color potential $V_C (L)$ can be explicitly isolated in
Coulomb gauge, Ref.~\cite{Christ:1980ku}, 
and it was shown in Ref.~\cite{Zwanziger:2002sh} that the
so-called ``Coulomb string tension'' $\sigma_C$ extracted from $V_C(L)$ is indeed
an upper bound to the Wilsonian string tension $\sigma$.\\
\noindent It is still a big challenge to understand the color confinement mechanism in
the continuum Yang-Mills theory, which requires to prove the area law for the
Wilson loop. In recent years several approaches to continuum Yang-Mills theory
have been intensively pursued. Among others there are the 
Dyson-Schwinger approach
in Landau gauge, Ref.~\cite{R5}, and more recently also in Coulomb gauge, Ref.~\cite{R6}, and a variational solution of the Yang-Mills Schr\"odinger equation in
Coulomb gauge, Ref.~\cite{Szczepaniak:2001rg,Feuchter:2004mk,Reinhardt:2004mm,Epple:2006hv,Epple:2007ut}. 
These approaches have given support for the
Gribov-Zwanziger confinement scenario, Ref.~\cite{R12,Zwanziger:1991ac}. 
What is, however, missing
in these approaches 
is the explicit non-perturbative 
evaluation of the Wilson loop, showing the emergence of the
area law. In the continuum theory the calculation of the Wilson loop is rendered
complicated due to the path ordering. In the context of supersymmetric
Yang-Mills theory a Dyson type of integral equation was proposed
 for the temporal
Wilson loop, which sums all planar ladder (or rainbow) diagrams, Ref.~\cite{Erickson:1999qv}.
This equation takes care of the path ordering, at least in an approximate
fashion, and was recently also applied to the temporal Wilson loop in ordinary
(non-supersymmetric) Yang-Mills theory, Ref.~\cite{Zayakin:2009jz}. Here 
the only input into this
Dyson equation  is the gluon propagator, which is
gauge dependent. 
In Ref.~\cite{Zayakin:2009jz} the gluon
propagator
was found by solving the Dyson-Schwinger equations of Yang-Mills theory in
Landau gauge in the rainbow-ladder approximation. In the present paper we will critically review the Dyson equation
for the Wilson loop, work
 out its ingredients and shortcomings, and apply it to Yang-Mills theory in Coulomb gauge. We study both the temporal and spatial Wilson loop.\\ 
\noindent The organization of the paper is as follows: In the next section we give a short heuristic derivation of the Dyson equation
for the Wilson loop proposed in Ref.~\cite{Erickson:1999qv} and critically
analyse the ingredients, the approximations involved and exhibit its range of
validity. In Sect.~III we solve this equation for the temporal Wilson loop
in Coulomb gauge assuming an instantaneous, temporal gluon propagator. 
In Sect.~IV the Dyson equation for the Wilson loop with arbitrary
gluon propagators is converted to a one-dimensional Schr\"odinger equation
 following Ref.~\cite{Erickson:1999qv}. In Sect.~V we use
this Schr\"odinger
 equation to calculate the spatial Wilson loop in the Hamiltonian
approach to Yang-Mills theory in Coulomb gauge. 
 Finally our conclusions are given in Sect.~VI.

\section{The Dyson equation for the Wilson loop}
\begin{figure}
\originalTeX
\centerline{
\includegraphics[height=6cm]{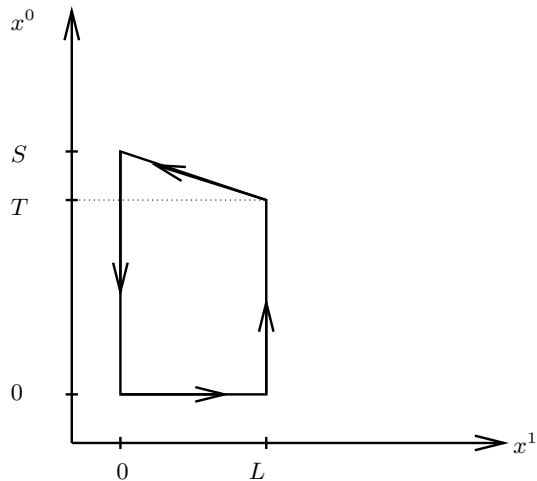}
\put(0,0){$x^1$}
\put(-100,-10){$L$}
\put(-150,-10){$0$}
\put(-190,20){$0$}
\put(-190,90){$T$}
\put(-190,110){$S$}
\put(-190,160){$x^0$}
}
\caption{\sl Trapezoidal temporal Wilson loop.}
\label{fig1}
\end{figure}

\begin{figure}
\originalTeX
\centerline{
\includegraphics[height=3cm]{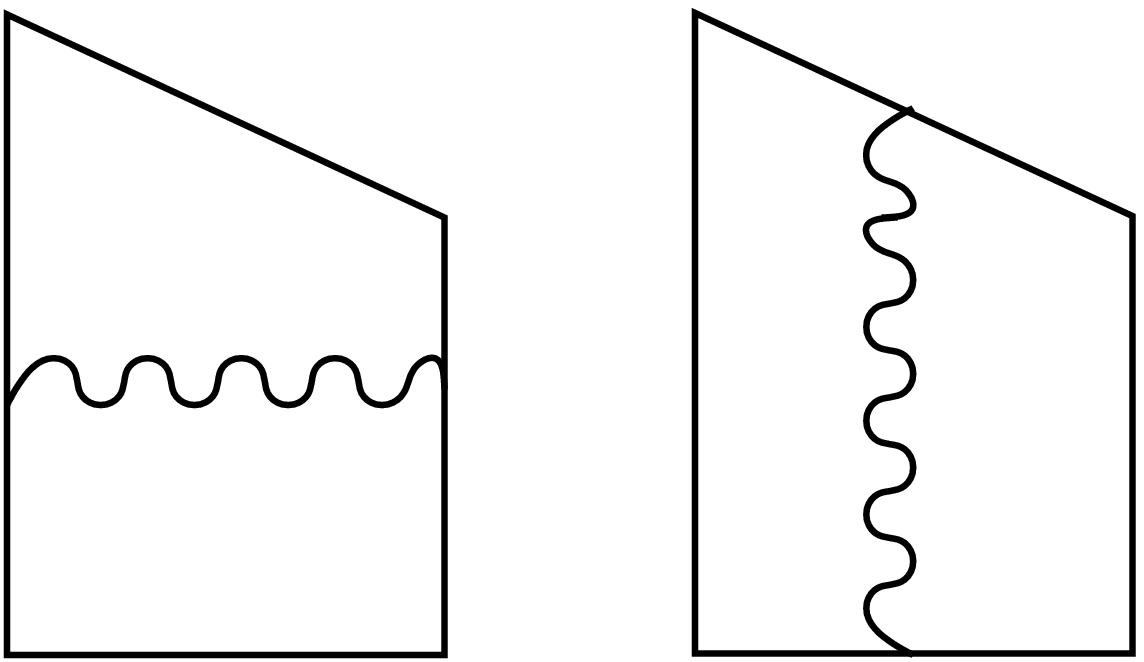}
\put(-35,-15){$(b)$}
\put(-125,-15){$(a)$}
}
\caption{\sl (a) Processes involved and (b) processes ignored in the Dyson equation (\ref{12}) illustrated in Fig.~\ref{fig3}. }
\label{fig2}
\end{figure}

Below we briefly sketch the derivation of the Dyson equation for the Wilson loop thereby exhibiting the approximations involved and working out its limitations.\\
\noindent Consider the Wilson loop  
 integral
\be
\label{1}
\langle W (\mathcal{C}) \rangle  = \left\langle \frac{1}{d_r} \mbox{tr} \, \mathcal{P} \, \exp{\left[- g \oint\limits_{\mathcal{C}} d x_\mu A_\mu (x) \right]}
\right\rangle \hk 
\ee
defined as expectation value via a functional
\be
\label{2b}
\langle \dots \rangle  =  \frac{\int D A_\mu \dots e^{- S [A]}}{\int D A_\mu
e^{- S [A]}} 
 \hk .
\ee
Here $\mathcal{C}$ is a closed loop in Euclidean space, $d_r$ is the dimension of the group representation and $\mathcal{P}$ denotes path ordering
along this loop. Furthermore $A_\mu = A^a_\mu T_a$ is the algebra valued gauge
field (referred to as gluon field) 
with $T_a$ being the anti-hermitean generators of the gauge group and $S [A]$ is the
action of the underlying gauge theory. In leading order 
perturbation theory only the
quadratic part of the action is kept
\be
\label{2}
S [A] = \frac{1}{2} \int 
d^d x \, d^d y \, A_\mu (x) D^{- 1}_{\mu \nu} (x, y) A_\nu(y) \; , \ee
with $D_{\mu \nu} (x, y)$ being the bare gluon propagator and $d$ being the number of space-time dimensions, and path
ordering can be ignored, yielding for the Wilson loop
\be
\label{3}
W (\mathcal{C}) = \exp \left[ - \frac{g^2}{2} C_2  I (\mathcal{C}) \right] \, = \, 1 - \frac{g^2}{2} C_2 I (\mathcal{C}) + \cdots \hk ,
\ee
where
\be
\label{4}
I (\mathcal{C}) = \oint\limits_{\mathcal{C}} d x_\mu \oint\limits_{\mathcal{C}} d y_\nu D_{\mu \nu} (x, y)
\ee
and
\be
\label{5}
T_a T_a = - \Id_{(d)} C_2
\ee
is the quadratic Casimir operator. For the gauge group $SU(N)$ we have
$C_2 = \frac{N^2-1}{2 N}$. 
Obviously, the perturbative gluon propagator $D (k) \sim \frac{1}{k^2}$ cannot
give rise to an area law.\\
Consider now 
a temporal planar trapezoidal Wilson loop $W = W (S, T; L)$ with two
parallel temporal sides of length $L$ and $T$, separated by a spatial distance
$L$, see Fig.~\ref{fig1}.
This loop consists of four straight paths. Accordingly, the loop integral
(\ref{4}) receives 16 contributions. In the limit $S, T \gg L$ the dominant
contribution comes from the two temporal paths (of length $T$ and $S$,
respectively). Keeping only these two paths and ignoring the contributions
where both integrals in (\ref{4}) run along the same temporal path, this loop
integral reduces to 
\be
\label{9}
I (\mathcal{C}) = - 2 \il^S_0 d x_0 \il^T_0 d y_0 D_{00} \lk \lk x_0 - y_0 \rk^2 + L^2
\rk \hk .
\ee
The diagram 
corresponding to this 
contribution to the Wilson loop is shown in Fig.~\ref{fig2}(a).
Since the gluon propagator (exact or perturbative) drops off for large distances
one may argue that for $S, T \gg L$ the contribution to the loop integral $I
(\mathcal{C})$ (\ref{4}) with gluon lines connecting the spatial paths as shown in Fig.~\ref{fig2}(b) are subleading compared to the contribution from the temporal
paths shown in Fig.~\ref{fig2}(a). Then in the standard way one may resum all
the ladder diagrams with gluon lines connecting the two
 temporal paths shown in
Fig.~\ref{fig3}(a) by a Dyson equation with kernel given by the gluon
propagator. This yields the integral equation originally proposed in the
context of supersymmetric gauge theories, Ref.~\cite{Erickson:1999qv}, 
which is illustrated in
Fig.~\ref{fig3}(b) and given by
\be
\label{12}
W (S, T; L) = 1 + g^2 C_2 \il^S_0 d s \il^T_0 d t D \lk \lk x (s) - s (t) 
\rk^2 \rk W (s, t; L)  \hk ,
\ee
where
\be
\label{13}
D \lk \lk x (s) - x (t) \rk^2 \rk = \dot{x}_\mu^- (s) D_{\mu \nu} \lk x (s), x 
(t) \rk
\dot{x}_\nu^+ (t)
\ee
and $x_\mu^\pm (s)$ denotes a parametrization of the two temporal
paths of the Wilson loop (see the Appendix for an explicit realization of these paths). 
Obviously, the summation of the ladder diagrams by the Dyson equation remains
valid when the full gluon propagator is used and does not lead to any
double counting.\\
\noindent The above given sketch of the derivation of the Dyson equation clearly exhibits
the limitations of this equation:
\begin{figure}
\originalTeX
\centerline{
\includegraphics[height=6cm]{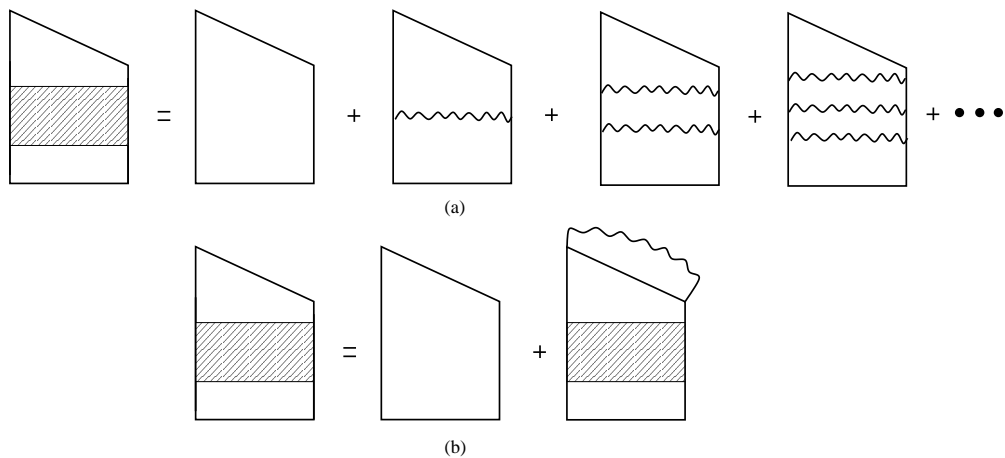}
}
\caption{\sl Graphical illustration to (a) the Dyson series for the Wilson loop and (b) the Dyson equation (\ref{12}).}
\label{fig3}
\end{figure}
\begin{enumerate}
\item The Dyson equation (\ref{12}) can be
only applied to strongly asymmetric loops consisting of two opposite long (above
temporal) and two opposite short (above spatial) paths. Otherwise it does not
make sense to include one pair of paths (Fig.~\ref{fig2}(a)) 
in the contour integration (\ref{4}) and
partially sum their ladder diagrams while neglecting the contour
 integrals of the
other pair of paths (Fig.~\ref{fig2}(b)) 
and their ladders. It is, however, not necessary that the
Wilson loop is  a temporal one. Eq. (\ref{12}) can be equally well
applied to strongly asymmetric spatial loops. Due to the limitation to 
asymmetric loops the static potential extracted from the solution of Eq.~(\ref{12}) will be accessible only for spatial distances $L \ll T$. In
particular, the limit $L \to \infty$ is not accessible by this equation. 
\item The solutions of the Dyson equation (\ref{12}) obviously 
satisfy the boundary
conditions
\be
\label{16}
W (S, T = 0; L) = 1 \hk , \hk W (S = 0, T; L) = 1 \hk .
\ee

\noindent However, for $T = 0$ or $S = 0$ the trapezoidal loop degenerates to a triangle
shaped loop, shown in  Fig.~\ref{fig4} (a)
 and the Wilson loop of the triangle shaped contour is
certainly not one. The wrong boundary value (\ref{16}) comes with no surprise since the limit $T=0$ or $S=0$ contradicts the assumption $S,T \gg L$ inherent in Eq.~(\ref{12}).
The wrong boundary value (\ref{16}) has consequences for the Wilsonian potential
extracted from the rectangular Wilson loop $W (S = T, T; L)$ shown in Fig.~\ref{fig4}(b). Assume the solution of the Dyson equation (\ref{12}) yields
asymptotically indeed an area law as expected for Yang-Mills theory. Due to the
boundary value (\ref{16}), $W (S = T, T; L)$ misses the contribution from half
of its enclosed area (i.e.~from the triangle shown in Fig.~\ref{fig4}(a)). If
the remaining triangle has to account for the full rectangle the string tension
has to be doubled. Consequently we expect from the solution of the Dyson
equation (\ref{12}) the double of the true string tension.
\bi
\begin{figure}
\originalTeX
\centerline{
\includegraphics[height=3cm]{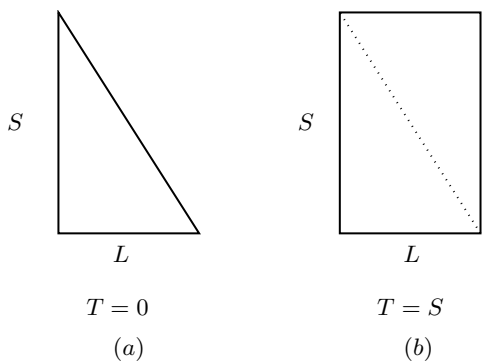}
\put(-30,-10){$L$}
\put(-140,-10){$L$}
\put(-70,40){$S$}
\put(-180,40){$S$}
\put(-40,-30){$T = S$}
\put(-150,-30){$T = 0$}
\put(-30,-45){$(b)$}
\put(-140,-45){$(a)$}
}
\caption{\sl (a) Triangle and (b) rectangular shaped loops representing the Wilson loop $W(S,T=0;L)$ and $W(S,S;L)$, see text.}
\label{fig4}
\end{figure}
\no
\item In the Dyson equation (\ref{12}) 
the only remnant
 from the representation of the gauge group is the quadratic Casimir (\ref{5}). Therefore, the Wilson loop
obtained as a solution of Eq.~(\ref{12}) will show strict Casimir scaling. It is
known, however, that Casimir scaling occurs only in the intermediate distance
regime, Ref.~\cite{Greensite:2003bk}. 
At large distances the string tension survives
only for the group representations with odd $N-$ality\footnote{$N$-ality
characterizes the 
representation of the center in an irreducible representation of
the group. The center of $SU(N)$ is $Z(N)$, consisting in the fundamental representation of the $N$ group elements
$z_n = \exp \lk i \frac{2 \pi n}{N} \rk \Id_{(N)} , n = 0, 1, 2, \dots, N - 1$.
Multiplication of a group element 
$g$ of an irreducible representation with $N$-ality
$k$ by the center element $z_n$ implies the multiplication of $g$ by the factor
$\exp \lk i \frac{2 \pi n}{N} k \rk$.} while the charges of the even $N-$ality 
representations are
screened. Thus again we find
that the Dyson equation (\ref{12}) can be appropriate only up to some
intermediate distances and is bound to fail for large (spatial) distances.
\item The Wilson loop, per se, is gauge
invariant while the right hand side of the Dyson equation (\ref{12}) is gauge
dependent through the gluon propagator. In fact, the gluon propagator is
non-vanishing only after the gauge has been fixed. Thus, Eq.~(\ref{12}) will
yield different results for the Wilson loop in different gauges. 
\item Finally, let us also mention that (except for the temporal Wilson loop in 
Coulomb gauge, see below)
the right hand side of the Dyson equation (\ref{12}) is not renormalization group invariant, while the
Wilson loop is.
\end{enumerate}
From the above analysis it is clear that the Dyson equation (\ref{12}) can be
applied only to strongly asymmetric loops with two different length scales and
that the smaller length scale has to be restricted from above to intermediate
distances. The limitations of the Dyson equation (\ref{12}) revealed above will
manisfest themselves in the applications of this equation to be given below. 

\section{The temporal Wilson loop in Coulomb gauge}
Since the Dyson equation (\ref{12}) for the Wilson loop is gauge dependent via the
gluon propagator it is preferable to use so-called ``physical'' gauges, for which
the gauge constraint and Gauss' law can be explicitly resolved, leaving a gauge
fixed theory of the physical degrees of freedom. Such a gauge is the Coulomb
gauge
\be
\boldsymbol{\partial} \cdot \vA = 0 \hk .
\ee
In addition this gauge has the advantage that 
the quantity $g A_0$ is renormalization
group invariant, which is not the case for covariant gauges like Landau gauge.
The Dyson equation (\ref{12}) for the temporal Wilson loop contains only the
temporal component of the gluon propagator
\be
g^2 D^{ab}_{00} (x, y) = \langle g A^a_0 (x) g A^b_0 (y) \rangle \hk.
\ee
Therefore in Coulomb gauge the Dyson equation (\ref{12}) is renormalization group
invariant and thus independent of the regularization and renormalization scheme.
Let us therefore investigate the Dyson equation in Coulomb gauge. In
Coulomb gauge the temporal gluon propagator has the structure
\cite{Cucchieri:2000hv}
\be
\label{7-16}
g^2 D^{ab}_{00} (x, y) = - \delta^{ab} V_C (|\vx - \vy|) \delta (x^0 - y^0) +
P^{a b}
(x, y) \hk .
\ee
Here $V_C (|\vx - \vy|)$ is the so-called non-Abelian Coulomb potential, which is
long-range and presumably
responsible for color confinement. It describes
 anti-screening of color charges, while $P (x, y)$, also being long-range,
describes ordinary screening. The latter
 is responsible for the breaking of the color
flux string between the external quarks when dynamical quarks are present. Both
terms have been calculated in perturbation theory
\cite{Drell:1981gu,Cucchieri:2000hv,Campagnari:2009km,Watson:2007mz}. The
non-Abelian Coulomb potential $V_C (|\vx - \vy|)$ was also non-perturbatively calculated
 both in
the Hamiltonian approach to continuum Yang-Mills theory
\cite{Feuchter:2004mk,Epple:2006hv} and on the lattice
\cite{Cucchieri:2002su,A21,B21}.
 At small distance it
behaves like the ordinary Coulomb potential $\sim 1/r$, while it rises linearly
at large distance $\sim \sigma_C r$ with a coefficient $\sigma_C$, referred to as
Coulomb string tension, which is somewhat larger than the Wilsonian string
tension $\sigma$ extracted from the (temporal) Wilson loop ($\sigma_C \sim
1.5 \sigma$ Ref.~\cite{B21}). \\
\noindent Since we are mainly 
interested in the infrared properties of the Wilson loop let us keep here the instantaneous (confinement relevant) 
part of the gluon propagator (\ref{7-16}) only. This forces us to consider 
rectangular
Wilson loops
\be 
\overline{W} (T; L) := W (T, T; L) \hk ,
\ee
for which the Dyson equation (\ref{12}) reduces  to
\be
\label{6-X1}
\overline{W} (T;L) = 1 - C_2 V_C (L) \il^T_0 d t \overline{W} (t;L) \hk .
\ee
This equation can be converted into the differential equation
\be
\label{10-18}
\frac{d}{d T} \overline{W} (T; L) = - C_2 V_C (L) \overline{W} (T, L)
\ee
with the boundary condition
\be
\label{10-19}
\overline{W} (T = 0; L) = 1 \hk ,
\ee
which is, contrary to the boundary condition of the general case (\ref{16}), indeed the correct boundary condition (the area enclosed by the loop vanishes for $T=0$).
The solution of Eqs.~(\ref{10-18}), (\ref{10-19}) or equivalently of 
(\ref{6-X1}) is given by
\be
\label{9-21}
\overline{W} (T; L) = \exp \lk - C_2 V_C (L) T \rk \hk .
\ee
Since the non-Abelian Coulomb potential rises linearly at large distances 
we have correctly obtained an area law, however, with the Wilsonian string
tension $\sigma$ 
replaced by the Coulomb string tension $\sigma_C > \sigma$. 
This is expected due to the
neglect of the non-instantaneous part $P (x, y)$ 
 of the gluon propagator (\ref{7-16}), 
which screens
the non-Abelian charge of the static quarks represented by the temporal Wilson
lines. Within the instantaneous approximation used for the gluon propagator
(\ref{7-16}) $\lk g^2 D_{00} (x) \to - V_C (\vx) \delta (x^0) \rk$ Eq.~(\ref{9-21}) is the correct result.
It is clear why in the present case the Dyson equation (\ref{12}) yields the
correct Wilson loop: The processes neglected in the Dyson equation (\ref{12}) (see Eq.~(\ref{2b})(b)) do not exist for an instantaneous gluon propagator.
\section{Extracting the static potential}
The charm of the Dyson equation (\ref{12}) for the Wilson loop is its
simplicity. It can be reduced to a one-dimensional Schr\"odinger equation and
from the corresponding 
ground state energy the static potential can be extracted 
\cite{Erickson:1999qv}.\\
\noindent Differentiation of Eq.~(\ref{12}) with respect to $S$ and $T$ yields the
differential equation
\be
\label{8x}
\frac{\partial^2 W (S, T; L)}{\partial S \partial T} = g^2 C_2 \, D \lk L^2 +
(S - T)^2 \rk W (S, T; L) \hk ,
 \ee
which together with the boundary condition (\ref{16}) is equivalent to the
integral equation (\ref{12}) but easier to solve. By introducing the variables
\be
\label{9x}
r  = \frac{S - T}{L} \hs , \hs R = \frac{S + T}{L} \hs , \hs
\frac{\partial^2}{\partial S \partial T}  = 
 \frac{1}{L^2} \lk
\frac{\partial^2}{\partial R^2} - \frac{\partial^2}{\partial r^2} \rk
\ee
this equation is separable \cite{Erickson:1999qv}. 
With the notation $W (R, r) := W (S, T; L)$ 
its solution can be expressed as 
\be
\label{10x}
W (R, r) = \sli_n  \varphi_n (r) \lk c^+_n 
\exp \lk \Omega_n R/2 \rk + c^-_n \exp \lk - \Omega_n R/2 \rk \rk
 \hk ,
\ee
where the $\varphi_n (x)$  satisfy the 1-dimensional Schr\"odinger equation
\be
\label{11x}
\left[ - \frac{d^2}{d r^2} + U (r) \right] \varphi_n (r) = -
\frac{\Omega^2_n}{4} \varphi_n (r) \hk 
\ee
with the potential
\be
\label{12x}
U (r) = - g^2 C_2 \, L^2 \, D \lk L^2 (1 + r^2) \rk \hk 
\ee
and the constants $c^\pm_n$ have to be chosen such that $W$ satisfies the
boundary conditions (\ref{16}), which in the variables (\ref{9x}) read $W (R, r
= R) = W (R, r = - R) = 1$ for all $R$. Since the potential is symmetric $U (- r) = U (r)$ the eigenfunctions 
$\varphi_n
(r), n = 0, 1, 2, \dots$ have definite
parity $\varphi_n (-r ) = (-)^n \varphi_n (r)$ and the boundery condition
implies that $c^\pm_n = 0$ for $n-$odd.
  By the symmetry of (\ref{10x}) the $\Omega_n$ can be chosen to be positive
 definite without loss of generality.\\
\noindent We are interested in the Wilson loop $W (S, T; L) \equiv W (R, r)$ for $L \ll S = T
 \to \infty$, i.e. $r = 0, R \to \infty$, which is related to the static
 potential by
 \be
 \label{10-28}
 V (L) = - \lim\limits_{T \to \infty} \frac{1}{T} \ln W (T, T; L) \hk .
 \ee
 In this limit and for $\Omega_n > 0$ the Wilson loop is dominated by the first
 term in (\ref{10x}) with the largest $\Omega_n$ for which $\varphi_n (0) \neq
  0$, which
 will usually correspond to the ground state $n = 0$ of (\ref{11x})
 \be
 \left. W (T, T; L) \right|_{T \gg L}
  = W (R \to \infty, r = 0) \approx \varphi_0 (0) e^{\Omega_0 R/2} \hk .
 \ee
 We thus obtain for the static potential (\ref{10-28})
 \be
 \label{F7X1}
 V (L) = - \frac{\Omega_0(L)}{L} + \text{const} \; . \hk 
 \ee
 Thus, to find the static Wilsonian potential we have to solve the Schr\"odinger equation (\ref{11x}) for
 its ground state as a function of $L$. As we have discussed above the validity of
 the Dyson-Schwinger equation (\ref{12}) and thus of the expression (\ref{F7X1})
 for the potential is restricted to not too large $L$. The precise range of
 validity will depend on the scale set by the gluon propagator which defines the
 potential $U(r;L)$ given in (\ref{12x}). This will be seen in the results.

 \section{Spatial Wilson loop in $D = 3 +1$ Coulomb gauge}
 At zero temperature, due to $O (4)$ symmetry, the Wilsonian potential can be
 extracted from either the spatial or temporal Wilson loops. In the Hamiltonian approach, the spatial loop is, however, more easily accessible than the temporal loop, once the vacuum wave functional has been
 determined. This is because the temporal Wilson loop requires also the time-evolution of the vacuum wave functional. As is clear from the discussion given in Sect.~2 the Dyson equation (\ref{12})
 can be equally well applied to spatial Wilson loops provided the loops are
 chosen asymmetrically with pairwise long and short paths. We are interested in
 the spatial Wilson loop in the Hamiltonian approach to Yang-Mills theory in
 Coulomb gauge \cite{Feuchter:2004mk,Reinhardt:2004mm,Epple:2006hv}. 
 In this case the expectation value is still given by Eq.~(\ref{2b}) where the functional integration runs now over the transversal spatial
 components of the gauge field only and the action $S [A]$ is defined by
 \be
 \label{spX1}
 e^{- S [A]} = J (A) |\psi [A]|^2 \hk .
 \ee
 Here $J (A)$ is the Faddeev-Popov determinant and $\psi [A]$ is the vacuum wave
 functional. In the approach of Ref.~\cite{Feuchter:2004mk} 
 the variational ansatz
 for the wave function $\psi [A]$ is such that the action functional defined by
 (\ref{spX1}) has the form
 \be
\label{11-32}
 S[A] = \int d^3 x d^3 y \, A^a_i (\vx) \omega (\vx, \vy) A^a_i (\vy) 
 \ee
 so that the static (spatial) gluon propagator is given by
 \be
 \label{spX2}
 D^{ab}_{ij} (\vx, \vy) = \langle A^a_i (\vx) A^b_j (\vy) \rangle = \delta^{ab}
 \frac{1}{2} t_{i j} (\vx) \omega^{- 1} (\vx, \vy) \hk .
 \ee

\begin{figure}
\originalTeX
\centerline{
\includegraphics[width=100mm]{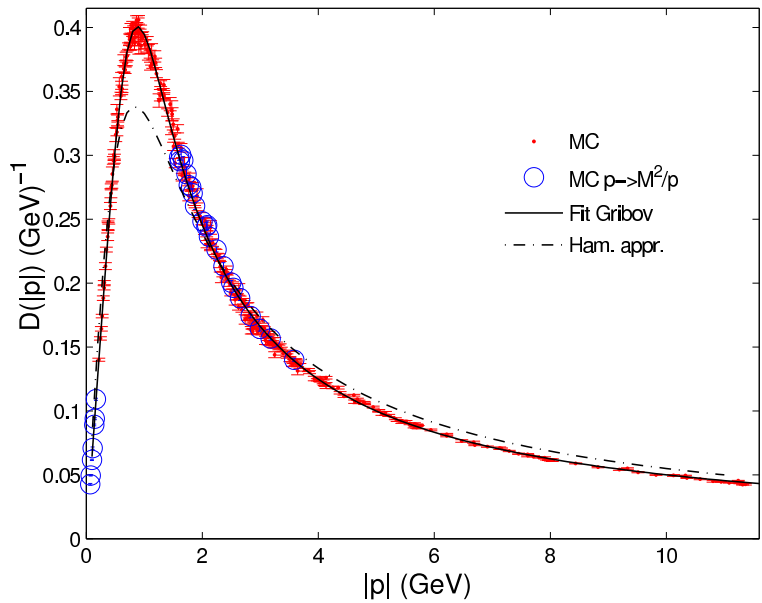}
}
\caption{\sl The gluon energy $\omega (k)$. Data points show the lattice results
of Ref.~\cite{Burgio:2008jr}. The full curve is the fit of the lattice data by
Gribov's formula (\ref{spX3}) while the dashed curve is
 the result of the variational
calculation \cite{Epple:2006hv}.}
\label{fig6}
\end{figure}

Here $\omega (\vx, \vy)$ is a variational kernel, which is found by minimizing
 the vacuum energy density, and $t_{ij}(\vx)$ is the projector to transversal gauge fields. 
 The result of the variational calculation is confirmed by a
 lattice calculation of the static gluon propagator \cite{Burgio:2008jr}, see Fig.~\ref{fig6}. In the infrared region, which is responsible for the string tension, 
the variational result almost perfectly matches the lattice data.  There
 are only small deviations in the intermediate momentum regime (see Fig.~\ref{fig6}), which shows some dependence on the single remaining undetermined
 renormalization parameter \cite{Reinhardt:2007wh}. Possibly, these deviations result partly from the neglect of the three-gluon vertex. 
One could perhaps fine-tune the remaining renormalization parameter
 to get a better agreement with the lattice data in the intermediate momentum
 regime. However, this region is irrelevant for the asymptotic behaviour of the
 Wilson loop. The lattice data for the kernel $\omega (k)$
 can be nicely
 fitted by Gribov's formula \cite{R12}
 \be
 \label{spX3}
 \omega (k) = \sqrt{k^2 + \frac{M^4}{k^2}} \hk , 
 \ee
 where $M$ is a mass scale which, in principle, is determined by the string
 tension. The
 fit to the lattice data yields \cite{Burgio:2008jr}
 \be
 \label{spX4}
 M \approx 880 \, \text{MeV} \approx 2 \sqrt{\sigma} \hk  .
 \ee
 We will use the static gluon propagator defined by Eqs.~(\ref{spX2}),
 (\ref{spX3}), (\ref{spX4}) in the Dyson equation (\ref{12}) (i.e. in the
 potential (\ref{12x}) of the Schr\"odinger equation (\ref{11x})) 
 to calculate the
 spatial Wilson loop in the Hamiltonian approach in Coulomb gauge.\\
\noindent When the expression (\ref{spX3}) is inserted for the (static) propagator $\overline{D} (k)
 = (2 \omega (k))^{- 1}$ the resulting integrals are
 UV-divergent (see Eq.~(\ref{F11XX}) in the Appendix). The reason is that Gribov's formula does not include the
 anomalous dimension of the gluon propagator. The logarithmic momentum
 dependence induced by the anomalous dimensions is difficult to see on the
 lattice. In the Hamiltonian approach to the continuum theory
 \cite{Szczepaniak:2001rg,Feuchter:2004mk}, which focuses on the infrared
 physics, the anomalous dimension escapes due to the particular 
 variational ansatz chosen for the
 vacuum wave functional, which leaves out the three-gluon vertex and thus the gluon loop contribution to the
 gluon self-energy. The latter is given in this approach by the ghost loop,
 which dominates in the infrared. The anomalous dimension of the static gluon 
 propagator was studied in Ref.~\cite{Schleifenbaum:2008ux}. Including the
 anomalous dimension the UV-term in Gribov's formula has to be modified to
\be 
\omega(k) = \sqrt{k^2 \, \lk 1 + a \ln^\gamma \frac{k}{M} \rk^2 \hk  + \frac{M^4}{k^2}} \label{anomalous}
\ee
 where the anomalous dimension (of the gluon propagator) is given by
 $\gamma = \frac{3}{11}$ \cite{Schleifenbaum:2008ux}. The parameter $a$ depends
 on $M$. Since the only effect of the anomalous dimension is to make the
 integrals (\ref{F11XX}) UV-convergent we will choose a small value of $a
 = 0.10$. The results do not depend on the precise value of the parameter $a$. Let us stress that the IR-behaviour of the gluon propagator is
 independent of the anomalous dimension $\gamma$.
 \begin{figure}
\originalTeX
\centerline{
\includegraphics[width=90mm,angle = 270]{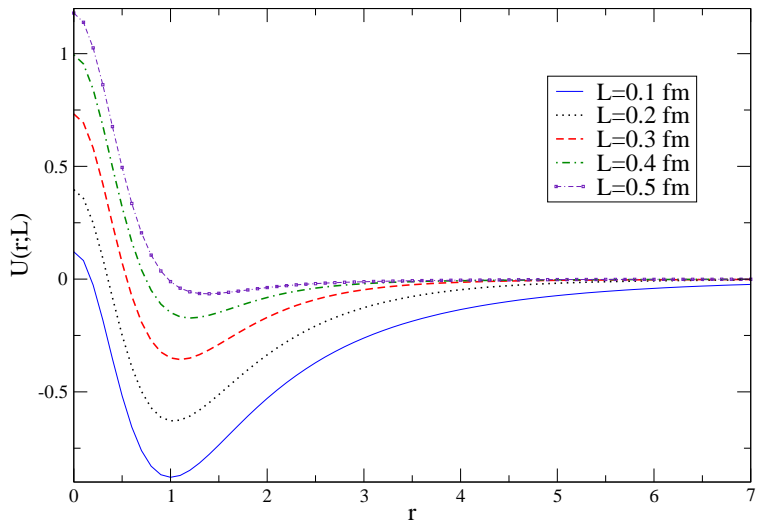}
}
\caption{\sl The potential $U(r;L)$ (\ref{12x}) for several distances $L$.}
\label{fig7}
\end{figure}
\no
 Fig. 7 shows the potential $U(r;L)$ (\ref{12x}) calculated from the static
 propagator (\ref{spX2}) in Coulomb gauge with $\omega (k)$ given by Eq.~(\ref{anomalous}) for various spatial distances $L$. $U (r; L)$ has the form of a double well centered at $r = 0$. The dip in the potential, necessary for the formation of a bound state, flattens as $L$ increases and vanishes for $L \geq 0.5$ fm. However, the bound state disappears already for $L \approx 0.35$ fm. The Wilson potential can then no longer be extracted from the ``ground state energy'' $\Omega_0 (L)$. This limits the use of the Schr\"odinger equation~(\ref{11x}) to rather small distances.\\
 \noindent The Wilsonian potential $V(L)$ obtained from the ground state eigenvalue $\Omega_0(L)$ via Eq.~(\ref{F7X1}) is shown in Fig.~\ref{fig8} as a function of
 the spatial distance $L$. The potential behaves like the ordinary Coulomb
 potential at small distances (below $0.15$ fm) and rises linearly in an
  intermediate distance regime between $0.15$ and $0.35$ fm.
At larger distances $L$ the Schr\" odinger equation (\ref{11x}) has no bound state and thus Eq.~(\ref{F7X1}) ceases to be applicable. The linear rise in the potential
is clearly exhibited when the perturbative potential $V_{\text{pert}}(L)$, calculated with
the perturbative gluon energy
\be
\omega_{\text{pert}}(k) = | \vec{k}| \lk 1 + a \ln^\gamma \frac{k}{M} \rk \hk , \label{perturbative}
\ee
is subtracted from the full potential, as shown in Fig.~\ref{fig8}. Except for the very
small distance regime, which is subject to numerical inaccuracies, the potential
void of its perturbative part can be nicely fitted by a linear function (see
Fig.~\ref{fig8})
\be
V (L) - V_{\text{pert}} (L) = c + \sigma L \hk , \label{scal-pert}
\ee
from which we extract the string tension
\be
\sigma \approx \left( 600 \, \text{MeV} \right)^2 \; . 
\ee
The string tension is about twice its input value of $\sigma = (440 \, \text{MeV})^2$ used
to fix the scale. This is in accord with our remark given in Sect. 2.
 \begin{figure}
\originalTeX
\centerline{
\includegraphics[width=70mm,angle=270]{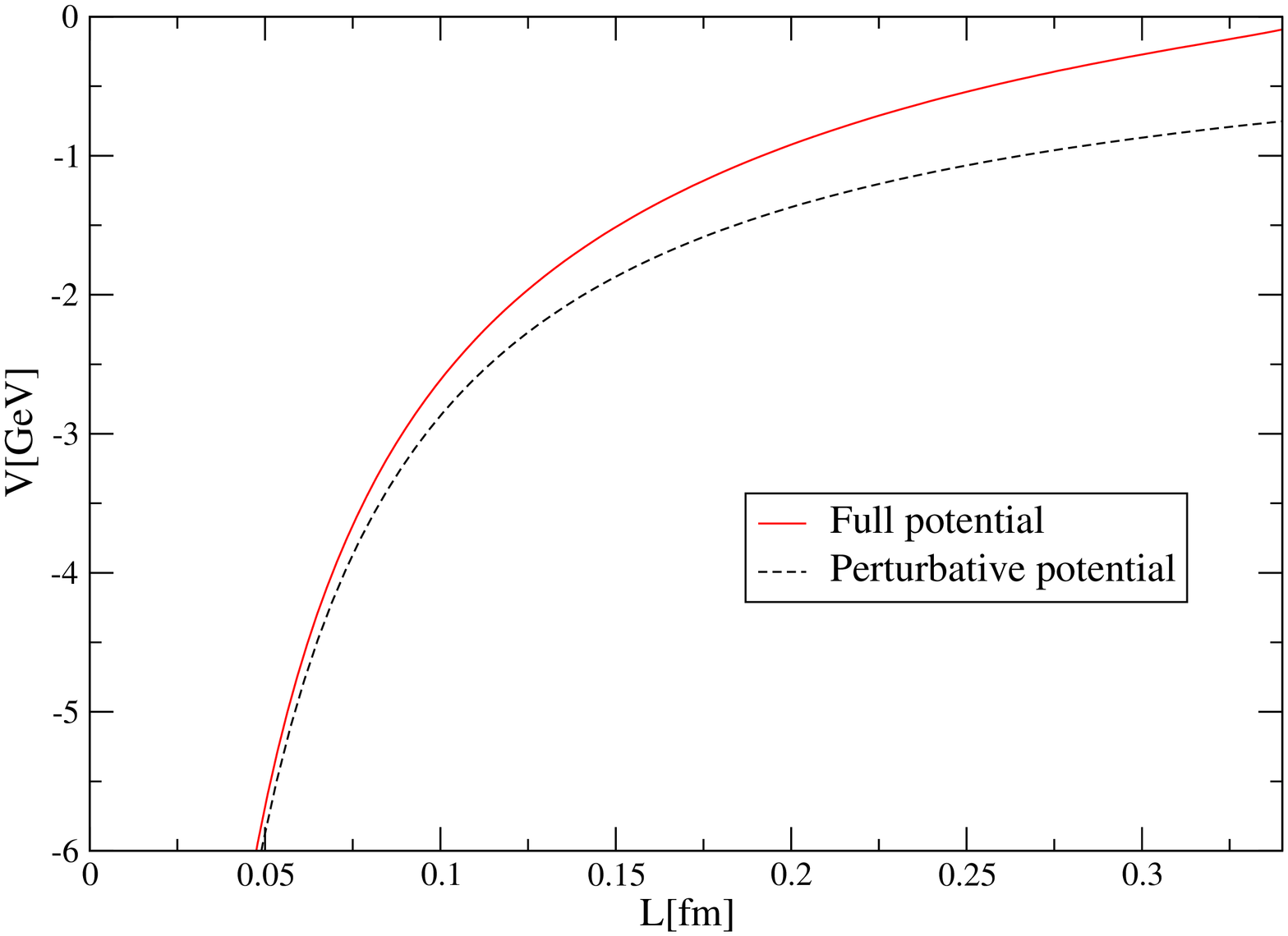}
\includegraphics[width=70mm,angle=270]{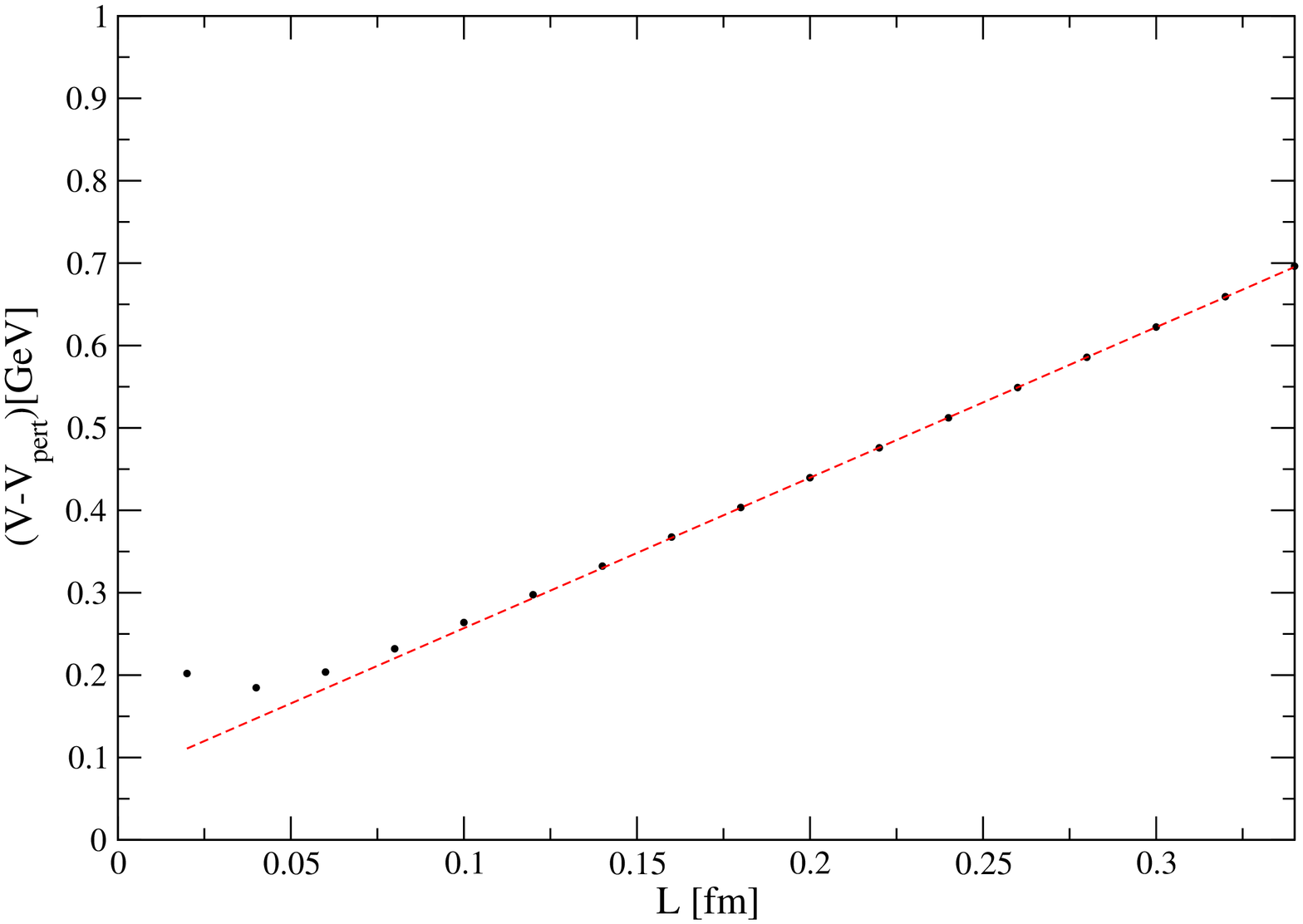}
}
\caption{\sl Left panel: The full static quark potential $V(L)$ obtained from the full propagator (\ref{anomalous}) and the perturbative potential $V_{\text{pert}}(L)$ obtained from the perturbative propagator (\ref{perturbative}). Right panel: The full potential minus its perturbative part.}
\label{fig8}
\end{figure}

 \section{Summary and Conclusions}
 We have studied the Dyson equation for the Wilson loop proposed in Ref.~\cite{Erickson:1999qv}
 in the context of supersymmetric gauge theory and applied in Ref.~\cite{Zayakin:2009jz} to
 ordinary Yang-Mills theory in Landau gauge. 
 We have examined the approximations involved and
 exhibited its range of applicability. We have shown that this equation can be
 applied only to strongly asymmetric Wilson loops with the smaller extension
 of the loop limited from above. This equation was then applied to the temporal Wilson loop in Coulomb gauge, keeping only the instantaneous part of the temporal gluon propagator. In this case the resulting Wilsonian potential coincides with the non-Abelian Coulomb potential. Finally, the Dyson equation was used to calculate the spatial Wilson loop in the Hamiltonian approach to Yang-Mills theory in Coulomb gauge. Although from the solution of this equation the Wilsonian potential can be extracted only up to intermediate distances, a strictly linearly rising potential was found when the perturbative potential was subtracted. The emergence of the area law in the Wilson loop
 found from the static Coulomb gauge propagator is consistent with the
 perimeter law for the 't Hooft loop found with the wave functional of the
 variational approach in~\cite{Reinhardt:2007wh}.\\
 \noindent An exact Dyson-Schwinger equation for the Wilson loop is given by the so-called
 loop equation \cite{25}. Like all Dyson-Schwinger equations this equation is not
 closed but is part of a tower of integral equations, which need truncations and
 approximations to arrive at a closed manageable set of equations. Such studies should eventually be carried out.

\begin{acknowledgments}
The authors are grateful to N. Brambilla, G. Burgio, D. Campagnari, M. Quandt, J. Rafelski, P.
 Watson and A. V. Zayakin for useful discussions. This work was supported by the Deutsche Forschungsgemeinschaft (DFG) under contract Re856/6-3 and by the Europ\" aisches Graduiertenkolleg ``Hadronen im Vakuum, Kernen und Sternen'' Basel-Graz-T\" ubingen.
\end{acknowledgments}
 
\appendix
\section{The Schr\"odinger potential}
 The potential~(\ref{12x}) in the 1-dimensional Schr\"odinger equation is, up to a constant prefactor, given by the quantity~(\ref{13}), which is the gluon propagator in coordinate space contracted with the two ``temporal'' paths $x^\pm (s)$. Below we will calculate this quantity for transversal gluon propagators in arbitrary dimensions $d$. \\
 \noindent In Euclidean coordinate space a (translationally invariant) transversal gluon propagator can be expressed by the Fourier integral
 \be
 \label{xyz2}
 D_{\mu \nu} \left( x \right) = \int \frac{d^d k}{\left( 2 \pi \right)^d} \left( \delta_{\mu \nu} - \hat{k}_{\mu} \hat{k}_{\nu} \right) e^{i k x} \bar{D} \left( k^2 \right)
 \ee
 where $\hat{k}_{\mu} = k_{\mu} / \sqrt{k^2}$ and $\bar{D} \left( k^2 \right)$ is a function of $k^2 = k_{\mu} k_{\mu}$ only. Obviously $D_{\mu \nu} \left( x \right)$ has the tensor structure
 \be
 \label{xyz3}
 D_{\mu \nu} \left( x \right) = \delta_{\mu \nu} I(x) - I_{\mu \nu}(x) 
 \ee
 where 
 \be
 \label{xyz4}
 I_{\mu \nu}(x) = \int \frac{d^d k}{\left( 2 \pi \right)^d} \, \hat{k}_{\mu} \hat{k}_{\nu} \, e^{i k x} \bar{D} \left( k^2 \right)
 \ee
 and $I(x) = I_{\mu \mu}(x)$. Since $x^\mu$ is the  only external vector, on which $I_{\mu \nu} (x)$ depends, the
 following tensor decomposition holds
 \be
 \label{F17Y2}
 I_{\mu \nu}(x) = \delta_{\mu \nu} I^1 (x) + \hat{x}_\mu \hat{x}_\nu I^2 (x) \hk
 , \hk \hat{x}_\mu = \frac{x_\mu }{\sqrt{x^2}} \hk .
 \ee
 Contracting this equation with $\delta_{\mu \nu}$ and $\hat{x}_\mu \hat{x}_\nu$
 we find
 \bea
 \label{F17X1}
 d I^1 (x) + I^2 (x) & = & I (x) \\
 \label{F17X2}
 I^1 (x) + I^2 (x) & = & \overline{I} (x) \hk ,
 \eea
 where $\overline{I} (x) = \hat{x}_\mu I_{\mu \nu} (x) \hat{x}_\nu \hk$. Solving the last two Eqs. for $I^{1, 2} (x)$ yields
 \bea
 (d - 1) I^1 (x) & = & I (x) - \overline{I} (x) \nonumber\\
 (d - 1) I^2 (x) & = & d \overline{I} (x) - I (x) \hk .
 \eea
 Inserting these relations into Eq.~(\ref{F17Y2}) we find for the transversal propagator in coordinate space
 \be
 (d - 1) D_{\mu \nu} (x) = \delta_{\mu \nu} \left[ (d - 2) I(x) + \overline{I}(x)
 \right] - \hat{x}_\mu \hat{x}_\nu [d \overline{I}(x) - I(x) ] \hk .
 \ee
  The
 remaining integrals 
 \be
 \label{F11XX}
 \begin{Bmatrix} I (x^2) \\ \overline{I} (x^2) \end{Bmatrix} =
 \int \frac{d^d k}{(2 \pi)^d} \, e^{i k \cdot
 x} \overline{D} (k^2) \, \begin{Bmatrix} 1 \\ (\hat{k} \cdot \hat{x})^2 \end{Bmatrix}
 \ee
 are worked out in the standard fashion using spherical
 coordinates in $k$-space and putting the $d$-axis of $k$-space parallel to
 $x_\mu$. The integrals over the first $d - 2$ angles are trivial yielding the
 volume of the unit sphere $S_{d - 2}$ in $d - 1$ dimensions
 \be
 \il_{S_{d - 2}} = \frac{(d - 1) \pi^{\frac{d - 1}{2}}}{\Gamma \lk \frac{d + 1}{2}
 \rk} \hk .
 \ee
 The integrals over the last angle yield with $z = \hat{x} \cdot \hat{k}$
 \bea
 \il^1_{- 1} d z \, e^{i z k x} & = & 2 \, \frac{\sin k x}{k x} \nonumber\\
 \il^1_{- 1} d z \, z^2 e^{i z k x} & = & 2 \, \frac{\sin k x}{k x} + \frac{4}{k^2
 x^2} \lk \cos k x - \frac{\sin kx }{k x} \rk \hk .
 \eea
 Inserting these results into (\ref{F11XX}) we obtain
 \bea
 \label{xyz10}
 I (x) & = & 2 \, C_d \il^\infty_0 d k k^{d - 1} \, \overline{D} (k^2) \, \frac{\sin k
 x}{k x} \nonumber\\
 \overline{I} (x) & = & 2 \, C_d \il^\infty_0 d k k^{d - 1} \, \overline{D} (k^2) \left[
 \frac{\sin k x}{k x} + \frac{2}{k^2 x^2} \lk \cos k x - \frac{\sin k x}{k x}
 \rk \right] \hk ,
 \eea
 where
 \be
 C_d = \frac{( d-1) \pi^{\frac{d - 1}{2}}}{(2 \pi)^d \Gamma \lk \frac{d +
 1}{2} \rk} \hk .
 \ee
 To work out the remaining one-dimensional integrals requires the explicit form of the scalar function $\bar{D} \left( k^2 \right)$. For definiteness we consider a planar temporal Wilson
loop in the 0-1-plane and choose the parametrization of the temporal pieces
 as
\be
\label{14}
x_\mu^\pm (s) = \lk s, \pm \frac{L}{2}, 0, \dots \rk \hk .
\ee
 For these paths we have
 \be
 \dot{x}_\mu^{\pm} (s) = (1, 0, 0, \dots ) =: \hat{e}_\mu \hk ,
 \ee
 and
 \be
 x_\mu := x_\mu^+ (S) - x_\mu^- (T) = L (r, 1, 0, 0, \dots)
 \ee
 so that
 \be
 \label{xyz15}
 x^2 = L^2 (1 + r^2)
 \ee
 and
 \be
 \lk \hat{x} \cdot \hat{e} \rk^2 = \frac{r^2}{1 + r^2} \hk .
 \ee 
 With these results we eventually find for the quantity~(\ref{13}) from~(\ref{xyz10})
 \bea
 \label{xyz12}
 \left( d - 1 \right) D \left( x^2 \right) =  
 \left[ \left( d - 2 \right) I \left( x^2 \right) + \bar{I} \left( x^2 \right) \right] - \frac{r^2}{1 + r^2} \left[ d \bar{I} \left( x^2 \right) - I \left( x^2 \right) \right] 
 \eea
 where $x^2$ is given by Eq.~(\ref{xyz15}).

\end{document}